\documentstyle[11pt]{article}
\newtheorem{theorem}{Theorem}[section]

\newenvironment{proof}{\begin{description}
                   \item[{\small {\bf Proof:}}] \small}{\hfill {\bf Q.E.D.}
                                                          \medskip
                                                       \end{description}}
\newtheorem{defi}{Definition}[section]
\newtheorem{prop}{Proposition}[section]
\newtheorem{lemma}{Lemma}[section]
\newtheorem{rem}{Remark}[section]
\newcommand{\bdef}{\begin{defi}}
\newcommand{\ede}{\end{defi}}
\newcommand{\bsat}{\begin{theorem}}
\newcommand{\esat}{\end{theorem}}
\newcommand{\bprop}{\begin{prop}}
\newcommand{\eprop}{\end{prop}}
\newcommand{\blem}{\begin{lemma}}
\newcommand{\elem}{\end{lemma}}
\newcommand{\brem}{\begin{rem}}
\newcommand{\erem}{\end{rem}}
\newcommand{\bbew}{\begin{proof}}
\newcommand{\ebew}{\end{proof}}
\newcommand{\be}{\begin{equation}}
\newcommand{\ee}{\end{equation}}
\newcommand{\bea}{\begin{eqnarray}}
\newcommand{\eea}{\end{eqnarray}}
\newcommand{\beas}{\begin{eqnarray*}}
\newcommand{\eeas}{\end{eqnarray*}}
\newcommand{\ben}{\begin{enumerate}}
\newcommand{\een}{\end{enumerate}}

\newcommand{\lra}{\longrightarrow}

\newcommand{\f}{\frac}

\newcommand{\Natural}{\mbox{I \hspace{-0.82em} N}}

\newcommand{\Real}{\mbox{I \hspace{-0.82em} R}}

\newcommand{\df}{\stackrel{\rm def}{=}}

\newcommand{\ng}{{\cal N}}

\newcommand{\gp}{\Lambda}

\newcommand{\lie}{{\cal L}}

\newcommand{\vesch}{ \Gamma(TM)}
\newcommand{\cvsch}{ \Gamma(T^\ast M)}
\newcommand{\bivsch}{\Gamma(\Lambda^2 TM )}

\newcommand{\la}{{\cal G}}

\newcommand{\glR}{{\cal GL}(n, \Real)}
\newcommand{\glRR}{{\cal GL}(2, \Real)}
\newcommand{\unit}{\mbox{1 \hspace{-0.95em} I}}
\newcommand{\semi}{\mathbin{\hbox{\hskip 2pt\vrule height 4.1pt 
                depth -.3pt width.25pt \hskip-2pt$\times$}}}

\begin{document}
\title{Linear Nijenhuis-Tensors and the Construction of Integrable Systems}
\author{Axel Winterhalder\\Fakult\"at f\"ur Physik der Universit\"at 
         Freiburg\\
        Hermann-Herder-Str. 3, 79104 Freiburg i. Br./ Germany}
        \date{August 1997\\Freiburg Preprint THEP 97/16}
\maketitle
\abstract
\noindent
A new method to construct Hamiltonian functions in involution is presented.  
We show that on left-symmetric algebras a Nijenhuis-tensor is given in a 
natural manner by the usual right-multiplication. Furthermore we prove 
that symplectic 
Lie-algebras carry the structure of a Poisson-Nijenhuis manifold. \\
{\it{keywords}}: Poisson-Nijenhuis structures, left-symmetric algebras, 
symplectic   Lie-algebras   

\section{Introduction}

A Poisson-Nijenhuis structure on a manifold (see \cite{KSM90}, \cite{KSM96}) 
provides a technique to construct a family of Hamiltonian functions in 
involution.
We first recall Poisson-Nijenhuis manifolds in general  
and then consider {\it{linear}} Poisson-Nijenhuis structures on a vector 
space. \-
It will be shown that linear Nijenhuis tensors on a vector space are in
one-to-one correspondence with  the structure of a so-called 
{\it{left-symmetric algebra}}. Such a structure
naturally exists on symplectic Lie algebras, i.e. Lie-algebras with a 
non-degenerate $2$-cocycle. 
Since on semi-simple Lie algebras a $2$-cocycle is always degenerate, we 
have to consider non semi-simple Lie algebras, which is
in contrast to usual constructions using semi-simple Lie algebras. 
Normally integrable systems are described
by the solutions of the modified Yang-Baxter equation. These are classified on
 semi-simple Lie algebras \cite{BD}. 
Left-symmetric algebras were first studied in \cite{VIN} and \cite{NIJ}.
It will be further seen that so-called {\it{symplectic Lie-algebras}} (see  
refs. \cite{LM}, \cite{DM}, \cite{BMO}) admit an interpretation as 
Poisson-Nijenhuis manifold in 
a natural manner. By taking the trace polynomials
of the linear Nijenhuis tensor, polynomial functions in involution can be 
constructed on these symplectic Lie algebras.\-      
The technique presented here is completely different from that of Mishchenko
and Fomenko (see ref. \cite{MF}).  
The Hamiltonian functions in involution on a symplectic Lie-algebra can be
pulled back in an appropriate way to its connected Lie-group to produce there 
Hamiltonian functions in involution.  

\section{Bihamiltonian Systems}

To motivate the notion of Poisson-Nijenhuis-manifolds we first introduce 
so-called bihamiltonian systems. These are dynamical sytems whose evolution 
in time is governed by two Hamiltonian functions.
To do this we need the notion of Poisson-bivectors.
\bdef
Given a manifold M, $ dim M = m $, then the tensor $ \gp ~ \in ~ \bivsch $  
is called Poisson-bivector, iff its Schouten-bracket vanishes, i.e.
\beas
\f{1}{2} [ \gp, \gp ](\alpha,\beta,\gamma)
 \df (\lie_{\gp^{\sharp}(\alpha)} \gp) (\beta,\gamma) 
   + d \alpha(\gp^{\sharp} (\beta),\gp^{\sharp}(\gamma)) = 0
\eeas
where $ \alpha, \beta, \gamma ~ \in ~ \cvsch $ and
$ \gp^{\sharp} : T^\ast M \longrightarrow TM $ is determined by: 
\beas 
  \alpha(\gp^{\sharp}(\beta)) = \gp( \alpha, \beta).
\eeas
$\lie$ denotes the usual Lie-derivative.
\ede

\brem A Poisson-bracket on $M$ is defined by: \-
$\{F,G\} \df \gp(dF,dG)$. Furthermore the Hamiltonian vector field of a
function $ f : M \lra \Real $ is given by: $ X_f \df \gp^{\sharp} df $.
The vanishing of the Schouten-bracket is equivalent
to the Jacobi-identity of $ \{, \} $. 
\erem
A {\it{bihamiltonian system}} is by definition given by two Hamiltonian 
functions 
$ H_1, H_2 : M \lra \Real $ and two Poisson-bivectors $\gp_1,~\gp_2$ such that:
\beas
       \gp_2^{\sharp}dH_2 \df X_{H_2} =  X_{H_1} \df \gp_1^{\sharp}dH_1 
\eeas
This can be read as the first part of the recursion relation:
\bea
       \gp_2^{\sharp}dH_{n+1} =  \gp_1^{\sharp}dH_n,~ n ~\in ~ \Natural
\eea
A simple consideration shows that
if the $\{H_n \}_{n ~\in~ \mbox{{\tiny{\Natural}}} } $ exist, they
are in involution with respect to each of the Poisson-brackets formed 
by $\gp_1 $ and $ \gp_2 $:
\beas
      \{H_n,H_m \}_1 = \{H_n, H_m \}_2 = 0,~ n,m ~ \in ~ \Natural 
\eeas
If we assume $ \gp_2^{\sharp} $ to be invertible, a necessary condition
for the existence of the sequence $\{H_n \}_{n \geq 3 } $ is $ d \alpha = 0$ 
with
$ \alpha \df \ng^{\ast} dH_1 $ and $ \ng^{\ast} \df 
(\gp_2^{\sharp})^{-1}\gp_1^{\sharp}$.
Hereby the mapping $\ng^{\ast}$ is the transpose of a
mapping $ \ng : TM \lra TM $: 
$(\ng^{\ast} \alpha)(X) \df \alpha( \ng X),~\alpha~\in~ \cvsch,~ X ~ \in ~ 
\vesch $. \-
The identity
\beas  
          d\alpha(X,Y) = - \f{1}{2} dH_1([\ng,\ng ](X,Y)),~ X,Y ~ \in ~  
\vesch,
\eeas
with the Nijenhuis torsion $[ \ng, \ng ]$ defined by:
\beas
      \f{1}{2} [ \ng, \ng ](X,Y) \df 
                  [\ng X,\ng Y] - \ng([\ng X,Y]+[X,\ng Y]) + {\ng}^2 [X,Y],
\eeas
(see e.g. \cite{KSM90}) implies $ d \alpha = 0$ for $[ \ng , \ng ] = 0 $. If  
$ \ng : TM \lra TM $ fulfills the
condition $ [ \ng , \ng ] = 0 $, it is called {\it{Nijenhuis tensor}}. 
Starting with a Nijenhuis tensor $\ng$ the vanishing   
of its Nijenhuis torsion implies the recursion relation: 
\begin{equation}  \label{Nrec}
      \ng^{\ast} dH_n = dH_{n+1},
\end{equation}
with the Hamiltonian functions $ H_n \df \f{1}{n} Tr {\ng}^n, ~ n \geq 1$,
 the trace polynomials of the Nijenhuis-tensor 
(see e.g. \cite{DW}). \\
We are now looking  for a Poisson-bivector $ \gp $ such that the tensor
$\gp_{\ng}$, defined by:
\beas
      \gp_{\ng}(\alpha,\beta) \df \gp( \alpha, \ng^{\ast} \beta)  
                           = \alpha(\gp^{\sharp} \ng^{\ast} \beta),~ \alpha, 
\beta~\in~ \cvsch, 
\eeas
is again a Poisson-bivector, i.e. $\gp_1 =  \gp_{\ng},~ \gp_2 = \gp$. \\ 
In this case we obtain the recursion relation:
\bea \label{rl}
      \gp_{\ng}^{\sharp} dH_n = \gp^{\sharp} dH_{n+1}.
\eea
This means, that the Hamiltonian functions 
$ \{ H_n \}_{n ~\in~ \mbox{{\tiny{\Natural}}} } $
are in involution with respect to the Poisson-bracket formed with 
$ \gp $ and $\gp_{\ng}$.
(See also \cite{NIJ}).
The conditions to be fulfilled by $\ng$ and $\gp$ such that $ \gp_{\ng} $ is a 
Poisson-bivector shall be examined in the following section.
\subsection{Poisson-Nijenhuis-Structures on Symplectic Manifolds}

In what follows we will restrict ourselves to the case where
the Poisson-bivector $ \gp $ is invertible. Then a symplectic form
$ \omega $ is defined on M by setting: 
$  \omega(X,Y) \df \gp( \gp^{\flat} X , \gp^{\flat} Y ),~
X,Y ~\in~ \vesch,~ \gp^{\flat} = (\gp^{\sharp})^{-1}  $ and $(M,\omega)$ is 
a symplectic manifold.\-      
We formulate the compatibility conditions for $\gp_{\ng}$ being 
a Poisson-bivector in the following theorem:                       
\bsat
Consider a symplectic manifold $(M,\omega)$ endowed with a mapping 
$ \ng: TM \lra TM $. Then  
the following holds:
\begin{quote}
    i) The antisymmetry of the tensor $\gp_{\ng}$ is equivalent to the 
symmetry of  
       $ \ng $ with respect to $ \omega $: $ \omega(\ng X, Y) = 
\omega(X, \ng Y) $. \\
                                   
       ii)Under the assumption that $ \ng $ is symmetric with respect to 
$ \omega $, \-   
       a $2$-form $F$ is defined by setting $F(X,Y) \df \omega(\ng X, Y)$.  
       Then the Schouten-bracket $[ \gp_{\ng}, \gp_{\ng} ]$ of  $ \gp_{\ng} $ 
fulfills the 
       identity:  
       \beas
       [ \gp_{\ng}, \gp_{\ng} ](\alpha, \beta, \gamma)
 & = &  dF(\gp^{\sharp}(\alpha), \gp^{\sharp}(\beta), \gp^{\sharp}(\gamma)) \\
& & - \omega([\ng,\ng](\gp^{\sharp}
  (\alpha),\gp^{\sharp}(\beta), \gp^{\sharp}(\gamma))
       \eeas
\end{quote}
\esat
The proof is straightforward and can be found in \cite{KSM96}.
\brem 
     Therefore, if $ \ng $ is symmetric with respect to $ \omega $ and in 
addition
     $ [ \ng, \ng ] = 0$ and $ dF = 0 $, then $\gp_{\ng}$ is 
a Poisson-bivector and
     we have the recursion relation (\ref{rl}). The trace polynomials are 
therefore
     in involution with respect to each Poisson-bracket defined by $\gp$ and 
$\gp_{\ng}$.\-
     If the compatibility conditions are fulfilled the triple 
$(M, \omega, \ng)$ 
     is called {\it{Poisson-Nijenhuis manifold}}. \\ 
     In \cite{NIJ} the compatibility conditions for $\omega$ and $\ng$ such 
that $(M, \omega, \ng)$ is a Poisson-Nijenhuis-manifold are formulated in 
a different but nevertheless equivalent manner. 
\erem
\section{Linear Nijenhuis-Tensors on Vector Spaces 
                 and Symplectic Lie-Algebras}

Up to now, only few Poisson-Nijenhuis structures are explicitely known 
(see e.g. \cite{DW}). To find new Poisson-Nijenhuis structures we now consider 
such structures on a vector-space V.\-
The simplest non-trivial choice is obviously a Nijenhuis-tensor depending
{\it{linearly}} on the co-ordinates. We therefore make the ansatz:
      \bea \label{ln}
           \ng(p)e_i = R_{ij}^k ~ x^j(p)~ e_k,
      \eea
where $ \{x^i \}_{i=1,\dots,n}, n = dim V$ are global coordinates on V with 
respect to
a basis $\{e_i \}_{i=1,\dots,n}$ and $\{ R_{ij}^k \}_{i,j,k = 1,\dots, n}$ are
constant coefficients. For $\ng$ defined above to be a Nijenhuis-tensor its 
Nijenhuis-torsion
has to vanish. Thus the coefficients $\{ R_{ij}^k \}_{i,j,k = 1,\dots, n}$ 
have to fulfill 
certain algebraic conditions as the following lemma shows:
      \blem
         The Nijenhuis-torsion of the tensor defined in (\ref{ln})
         has the coordinate expression:
           \bea
               \f{1}{2} {[ \ng , \ng ]_{ij}^k}(p) = 
                   (-R_{ml}^k (R_{ij}^m - R_{ji}^m) - 
                         (R_{il}^m R_{jm}^k - R_{jl}^m R_{im}^k)) {x^l}(p)
           \eea                   
     \elem
This equation admits a surprising interpretation. To this purpose we interprete
the coefficients $\{ R_{ij}^k \}_{i,j,k = 1,\dots, n}$ as structure constants 
of a multiplication on V making it to an algebra by setting:
           \bea 
                e_i \cdot e_j = R_{ij}^k e_k .
           \eea     
If we furthermore define the associator of this algebra as follows:
    \bea
         [x,y,z] \df (x \cdot y) \cdot z - x \cdot ( y \cdot z)
    \eea
then the Nijenhuis-torsion has an elegant expression as the subsequent theorem
shows:          
   \bsat
Given the tensor defined in formula (\ref{ln}), then its Nijenhuis-torsion 
$ [ \ng, \ng ] $ fulfills the relation:
\bea  \label{arel}
       [ \ng, \ng ]_{(p)}(x,y) = [x,y,p]-[y,x,p], ~ p,x,y~ \in ~  V 
\eea
\esat
\brem
Because of formula (\ref{arel}) linear Nijenhuis-tensors on a vector space 
are in one-to-one correspondence with
left-symmetric multiplication structures on this vector space. 
Furthermore each linear Nijenhuis-tensor is given 
by the right-multiplication: $ \ng_{p}(x) = x \cdot p = R_p(x) $.
\erem
Thus the identity:
     \bea \label{ls}
           [x,y,z] = [y,x,z]
     \eea
has to be fulfilled for all $ x,y,z \in V $ in order the Nijenhuis-torsion of 
$\ng$
to vanish.
Since the associator measures the lack of associativity of an algebra the 
algebra structure above is in general {\it{non-associative}}. Algebras whose
 associator fulfills
(\ref{ls}) are called {\it{left-symmetric algebras}}. They will be introduced
in the following section.

\subsection{Left-Symmetric Algebras}

Left-symmetric algebras appeared first in \cite{VIN} and \cite{NIJ} and 
are also
called {\it{Koszul-Vinberg algebras}}. Their algebraic structure is 
studied in \cite{HEL71} and \cite{HEL79}. \\ 
Consider an algebra $\cal{A}$ and define an associator on $\cal A$ as above. 
\bdef
$ \cal{A} $ is called  
{\it left-symmetric} iff for all $ x,y,z ~ \in ~ \cal{A} $ the identity:
\beas
         [x,y,z] = [y,x,z],
\eeas          
i.e.
\beas          
          (x \cdot y) \cdot z  - x \cdot (y \cdot z) =  
                                 (y \cdot x) \cdot z  - y \cdot (x \cdot z)
\eeas
holds.
\ede
Thus, as mentioned above, left-symmetric algebras are
in general non-associative, whereas associative algebras are trivial examples 
for left-symmetric algebras. \\
Nevertheless, by setting $ [x,y] \df x \cdot y - y \cdot x $ 
a Lie-bracket is defined. 
The Jacobi-identity follows because of the left-symmetry property of the 
associator:
\beas
        [x,[y,z]]+[y,[z,x]]+[z,[x,y]] &=&  [y,x,z] - [x,y,z] \\
                                      &+&  [z,y,x] - [y,z,x] \\
                                      &+&  [x,z,y] - [z,x,y] \\               
                                      &=&   0
\eeas
Therefore every left-symmetric algebra gives rise to a Lie-algebra.\\
\brem
The geometric interpretation of a left-symmetric multiplication is given by
a left-invariant flat torsion-free connection on the connected Lie-group
$G_{\cal A}$ of $\cal A$ (see e.g. \cite{HEL71}).  
\erem
\subsection{Symplectic Lie-Algebras}

Symplectic Lie-algebras are studied in \cite{LM}, \cite{DM} and \cite{BMO}. \\
A Lie-algebra $\cal{G}$ endowed with a non-degenerate $2$-cocycle 
{$ \omega: \cal G  \times \cal G \lra \Real $}, i.e. which fulfills the 
cocycle identity:
\beas \label{cocl}
       \omega([x,y],z) + \omega([y,z],x) + \omega([z,x],y) = 0
\eeas
is called a {\it{symplectic Lie-algebra}}. \\
By 
\begin{equation} \label{lsym} 
 \omega( x \cdot y, z) \df - \omega(y,[x,z]) 
\end{equation}
a left-symmetric multiplication is defined on $ \cal G $. 
\bbew
The left-symmetry property $ [x,y,z] = [y,x,z] $ is equivalent to the fact 
that the 
left-multiplication $ L_x(y) = x \cdot y $ fulfills the representation property
\beas
      L_{[x,y]} = L_x L_y - L_y L_x.
\eeas
With (\ref{lsym}) one has: 
$ L_x(y) = \omega^{\sharp}(ad(x). \omega^{\flat}(y)) $, where
$ \omega^{\flat} : \cal G \lra {\cal G}^{\ast} $ is defined by:
$ (\omega^{\flat}(x))(y) \df \omega(x,y)$, with $~x,y~\in~{\cal G} $, \- 
$ \omega^{\sharp} = (\omega^{\flat})^{-1}$  
and $ ad $ denotes the adjoint representation of $ \cal G $ on itself.
With these considerations the representation property of the 
left-multiplication
is a direct consequence of the representation property of the adjoint 
representation.
\ebew
We now consider $ \omega $ as constant symplectic form on $ \cal G $.
As it will be shown, the Nijenhuis-tensor $ \ng_{(p)} = R_p $ is symmetric 
with respect to $ \omega $. Further the exterior differential of 
the 2-form $F$ formed
with $ N_{(p)} = R_p $ and $ \omega $ vanishes:
\bsat
Consider a symplectic Lie-algebra $ (\cal G, \omega) $. Then the triple 
$({\cal G}, {\omega}, R_p)$ is a Poisson-Nijenhuis manifold. 
\esat
\bbew
First we observe that the Nijenhuis-tensor is symmetric with respect to 
$\omega$:
\beas
\omega(R_p(y),z) = \omega(y,R_p(z)). 
\eeas
Further one has:
\beas
dF_{(p)}(x,y,z) &=&
                     \omega(y \cdot x, z) - \omega(x \cdot y, z)
                    + \omega(x \cdot z, y) \\
                &=& - \omega(x,[y , z]) - \omega(y,[z , x])  
                    - \omega(z,[x , y]) \\               
                &=& 0,
\eeas
since $ \omega $ fulfills the cocycle-identity (\ref{cocl}). 
\ebew
\brem
     A simple argumentation shows that semi-simple Lie-algebras never admit 
a non-degenerate $2$-cocycle. 
\erem

\subsubsection{Explicit Expressions}

To give an explicit expression for the Hamiltonian functions $H_n$, it 
is necessary 
to make some considerations before. 
Given an arbitrary left-symmetric algebra $ \cal A $.
If the left- respectively the right-multiplication is defined by setting 
$L_x y \df x \cdot y $ respectively 
$ R_x y \df y \cdot x, ~ x,y ~ \in ~ \cal A $, then 
the left-symmetry property can be rewritten as follows:
\bea \label{idR}
          R_x R_y - R_{y \cdot x} = R_x L_y - L_y R_x, ~ x,y ~ \in ~ \cal A
\eea
Defining the linear functional $\tau : \cal A \lra \Real$ by
${\tau}(x) \df TrR_x $ we obtain:
\beas 
  {H_n} (x) &=& \f{1}{n} ~ Tr(R_x)^n \\  
           &=& \f{1}{n} ~ \tau((R_x)^n x),  
\eeas
(see also \cite{HEL79}).
Further a symmetric bilinear form is defined by:
\beas
        b(x,y) \df Tr ~ R_x R_y = Tr~ R_{y \cdot x}
\eeas
Thus for example the Hamiltonian function $H_2$ can be expressed as follows:
\beas
      {H_2} (x) = \f{1}{2}~b(x,x),~ x \in ~  \cal A
\eeas
With identity (\ref{Nrec}) a formula for the differential
$dH_n$ is obtained:
\beas
    dH_n(x)(h) =  {\tau}((R_x)^{n-1}h),
\eeas
where $x,h~\in~ {\cal A}$.
\brem
In the case of a symplectic Lie-algebra $ ( \la, \omega) $ the identity: 
$ \tau (x) = -2~Tr~ad(x) $ holds. Therefore, if $ \la $ is unimodular, the 
trace polynomials 
$ {\{ H_n \}}_{n ~ \in~ \mbox{{\tiny{\Natural}}}} $ all vanish 
(see e.g. \cite{LM}).
\erem

\subsection{Example}

As an example we consider the semidirect product $\glR \semi {\Real}^n$, where
$\glR$ denotes the Lie-algebra of real $n \times n$-matrices. \\
To obtain a symplectic form we define $\omega(x,y) \df \nu([x,y])$, where 
{$\nu : \glR \times {\Real}^n \lra \Real $} is chosen 
such that $\omega$ is invertible (see \cite{BMO} for further details).\\
On $ \glR \times {\Real}^n $ a Lie-bracket is defined as follows:
 \beas
     [(A,x),(B,y) ] = (AB - BA, Ay - Bx),
 \eeas     
where $A,B~\in~ \glR$ and $x,y ~ \in ~ \Real^n$.
Defining the $1$-form $\nu$ by:
  \beas   
     \nu (A,x) \df Tr(MA) + g(x),~g~\in~{{\Real}^n}^{\ast},~M~\in~\glR
  \eeas   
we have further:
  \beas
       \omega((A,x),(B,y)) = g(Ay) - g(Bx) + Tr([M,A],B). 
  \eeas
In general the explicit expressions in co-ordinates of the Hamiltonian
functions $H_n$ are rather complicated. Thus we restrict ourselves to the
case $n = 2$: \\
To obtain a basis for $\glRR \semi {\Real}^2$ we make the decomposition 
$ \glRR = {\cal SL}(2, \Real) \oplus \{ \unit \} $, where 
$ {\cal SL}(2, \Real) $
is the Lie-algebra of traceless real $ 2 \times 2 $-matrices and $ \unit $ 
is the
unit-matrix of $ \glRR $. 
With the basis
 \beas
     H = \left( \begin{array}{cc}
                1 & 0 \\
                0 & -1 
                \end{array} \right ) ~~
     X_+ = \left( \begin{array}{cc}
                0 & 1 \\
                0 & 0 
                \end{array} \right ) ~~
     X_- = \left( \begin{array}{cc}
                0 & 0 \\
                1 & 0 
                \end{array} \right )
 \eeas                 
of $ {\cal SL}(2, \Real)$ and the basis
 \beas
    e_1 = \left( \begin{array}{c}
                1  \\
                0  
                \end{array} \right ) ~~
    e_2 = \left( \begin{array}{c}
                0 \\
                1 
                \end{array} \right ) ~~
  \eeas
of $ \Real^2 $, a basis for $\glRR \semi {\Real}^2$ is obtained:
  \beas
       && v_1 = (0,e_1), ~~ v_2 = (0,e_2), ~~ v_3 = (\unit , 0), ~~ \\
       && v_4 = (H , 0), ~~ v_5 = (X_+ , 0), ~~  v_6 = (X_- , 0).
   \eeas                                   
We choose 
   \beas
         M = \left( \begin{array}{cc}
                l & 1 \\
                0 & l 
                \end{array} \right )
   \eeas
and $ g(x) = a < e_1, x > $, where $x~\in~{\Real}^n,~ a,l~\in~\Real$ and 
$ < ~ , ~ > $ 
is the usual scalar product in $ \Real^n $.   
If $ \{ {\bar x}_i \}_{i=1, \dots 6} $ denote the co-ordinates with respect to
the basis above, we make the change of co-ordinates:
  \beas
       && {\bar x}_1 = x_1 - x_5 + \f{2}{a} x_6, ~~
          {\bar x}_2 = x_2, ~~
          {\bar x}_3 = x_3 - x_4, ~~ \\
       && {\bar x}_4 = x_4 ~~
          {\bar x}_5 = x_1 + x_5, ~~
          {\bar x}_6 = x_6, 
  \eeas     
such that in these co-ordinates $H_2(x) = \f{1}{2} b(x,x), x~\in~\glRR $
has got its standard form, i.e. the bilinear form b is diagonalized. Then
we obtain finally: 
      \beas
            H_1(x) &=& - 4 x_3 + 4 x_4 \\
            2H_2(x) &=& -4a {x_1}^2 + 4 {x_3}^2 + 8 {x_4}^2 + 4a {x_5}^2 \\
            3H_3(x) &=& -4 {x_3}^3 + 16 {x_4}^3 
                        + 6a(x_1 - x_5)(x_1 + x_5)(x_3-2x_4) \\
                        & & - 6a{x_2}(x_1 + x_5)^2                      
      \eeas        
It can easily be seen that $dH_1 \wedge dH_2 \wedge dH_3 \neq 0 $ almost 
everywhere.
Therefore the functions $ H_1,H_2,H_3 $ form a complete set of 
Hamiltonian functions
 in involution on $ \glRR \semi \Real^2 $. 
   \brem
        The family of functions in involution 
${\{H_n\}}_{n ~ \in ~ \mbox{\tiny{\Natural}}}$ 
        on the Lie-algebra $\la$ may be appropriately 
        pulled back to the connected Lie-group G of $\la$. On G 
        a symplectic form is defined by pulling back the symplectic form 
$\omega$ on $\la$. 
        The familiy of functions on G obtained above are then again 
in involution with respect to the 
Poisson-bracket which is given by this symplectic form on G.      
   \erem

\section{Conclusion}

The functional independence of the the trace polynomials can be proven 
up to now only analytically and for the simplest cases. It 
is still to be examined, whether the
functional independence can be proved also algebraically, i.e. 
by using the algebraic 
structure of left-symmetric algebras (see \cite{HEL79}).\-  
This would be in analogy to the proof of the functional independence of the
              Mishchenko-Fomenko polynomials (see \cite{MF}). \\
              On the connected Lie-group G
              of a symplectic Lie-algebra, a canonical momentum mapping
              $J: G \lra \la^{\ast}$ exists. The Hamiltonian functions in 
involution on 
              $\la$ described above may be pulled back appropriately on 
$\la^{\ast}$
              and from there as already mentioned 
              with J to the Lie-group G producing Hamiltonian functions 
              in involution. \\
              In \cite{BMO} the semidirect product
              $ \glR \semi \Real^n $ is considered 
              as symplectic Lie-algebra.\- 
              Furthermore a Poisson-morphism between the connected Lie-group
              $ GL(n,\Real) \semi \Real^n$ belonging to 
$ \glR \semi \Real^n $ and
              and the cotangent bundle of the configuration space of the 
translating
              top, $T^{\ast} (SO(n) \semi \Real^n )$ is constructed. By 
pulling back the Hamiltonian
              functions in involution on $ GL(n,\Real) \semi \Real^n$ via this
              Poisson-morphism there is a possible physical interpretation
              for the so-obtained Hamiltonian functions in involution.

\end{document}